%% file: template.tex
\documentclass{vgtc}                
\ifpdf
  \pdfoutput=1\relax                   
  \usepackage{graphicx}                
  \DeclareGraphicsExtensions{.pdf,.png,.jpg,.jpeg} 
\else
  \usepackage{graphicx}                
  \DeclareGraphicsExtensions{.eps}     
\fi%

\graphicspath{{figures/}{pictures/}{images/}{./}} 

\usepackage{microtype}                 
\PassOptionsToPackage{warn}{textcomp}  
\usepackage{textcomp}                  
\usepackage{mathptmx}                  
\usepackage{times}                     
\usepackage{cite}                      
\usepackage{tabu}                      
\usepackage{booktabs}                  
\usepackage[dvipsnames]{xcolor}
\usepackage{xspace}
\usepackage{comment}
\usepackage{enumitem}
\usepackage{tablefootnote}
\usepackage{caption}
\newcommand{\tool}[0]{\textsc{Argo Scholar}\xspace{}}
\newcommand{\participants}[0]{122}

\onlineid{0}

\vgtccategory{Research}
\vgtcpapertype{System}

\renewcommand\footnotemark{}

\title{Evaluation of \tool{} with Observational Study}

\newcommand{\authorgap}{\hspace{10pt}}
\author{
  Kevin Li\textsuperscript{\textrm 1} %
  \thanks{\textsuperscript{\textrm 1}Georgia Institute of Technology. \{\href{mailto:kevin.li@gatech.edu}{kevin.li}$\mid$\href{mailto:alexanderyang@gatech.edu}{alexanderyang}$\mid$\href{mailto:emontoya30@gatech.edu}{emontoya30}$\mid$
  \href{mailto:aupadhayay3@gatech.edu}{aupadhayay3}$\mid$\href{mailto:zzhou406@gatech.edu}{zzhou406}$\mid$\href{mailto:jonsaadfalcon@gatech.edu}{jonsaadfalcon}$\mid$\href{mailto:polo@gatech.edu}{polo}\}@gatech.edu} \authorgap
  Haoyang Yang\textsuperscript{\textrm 1}
  \authorgap
  Evan Montoya\textsuperscript{\textrm 1}
  \authorgap
  Anish Upadhayay\textsuperscript{\textrm 1}
  \authorgap \\
  Zhiyan Zhou\textsuperscript{\textrm 1}
  \authorgap
  Jon Saad-Falcon\textsuperscript{\textrm 1}
  \authorgap
  Duen Horng (Polo) Chau\textsuperscript{\textrm 1}
  \authorgap
}

\input{txt/Section-Abstract}

\CCScatlist{ 
 \CCScatTwelve{Human-centered computing}{Visu\-al\-iza\-tion} {Visualization design and evaluation methods}{}
}

\vgtcinsertpkg
\definecolor{linkColor}{RGB}{6,125,233}

\begin{document}


\input{txt/Section-Intro}
\input{txt/Section-Background}
\input{txt/Section-Eval}
\input{txt/Section-Conclusion}



\bibliographystyle{abbrv-doi}
\bibliography{template}
\end{document}

%% file: txt/Section-Abstract.tex
\abstract{
Discovering and making sense of relevant literature is fundamental in any scientific field.
Node-link diagram-based visualization tools can aid this process; 
however, existing tools have been evaluated only on small scales.
This paper evaluates \tool{}, an open-source visualization tool designed for interactive exploration of literature and easy sharing of exploration results. 
A large-scale user study of \participants{} participants from diverse backgrounds and experiences showed that
\tool{} is effective at helping users find related work and understand paper connections, and incremental graph-based exploration is effective across diverse disciplines. 
Based on the user study and user feedback, we provide design considerations and feature suggestions for future work.
}

%% file: txt/Section-Intro.tex
\firstsection{Introduction}

\maketitle
Discovering and making sense of relevant literature is fundamental
in any scientific field.  
Academic search engines are common starting point; however, keyword searching does not provide users with all relevant papers, so incremental exploration, a common and effective method where users iterate from familiar (identified via search engines) to novel works while filtering based on paper attributions, is also employed \cite{Snyder2019LiteratureRA}. 
Visualization can aid this process,
such as node-link diagram-based tools that support the visual exploration of citation networks with a bottom-up approach that allows users to grow their own networks and understanding of its papers \cite{Chau2011ApoloMS}. 
However, the evaluations of existing tools often only consist of a dozen or so participants for certain selected domains \cite{Chau2011ApoloMS, Sultanum2020UnderstandingAS}. A  large-scale evaluation of such node-link based literature exploration tools across diverse domains could generate different discoveries and insights.

This paper presents a follow-up, large-scale evaluation of \tool{}\cite{Li2021ArgoSI}, open-sourced, web-based visualization tool that facilitates the incremental exploration of literature to gain insight and uncover relevant work.
We were interested to see how receptive participants from diverse backgrounds were to node-link diagram representations of citation networks and what features of literature exploration tools did participants find the most helpful.
To answer these questions, we recruited \participants{} participants from a wide range of backgrounds and research experiences.

\begin{figure}
  \includegraphics[width=\linewidth]{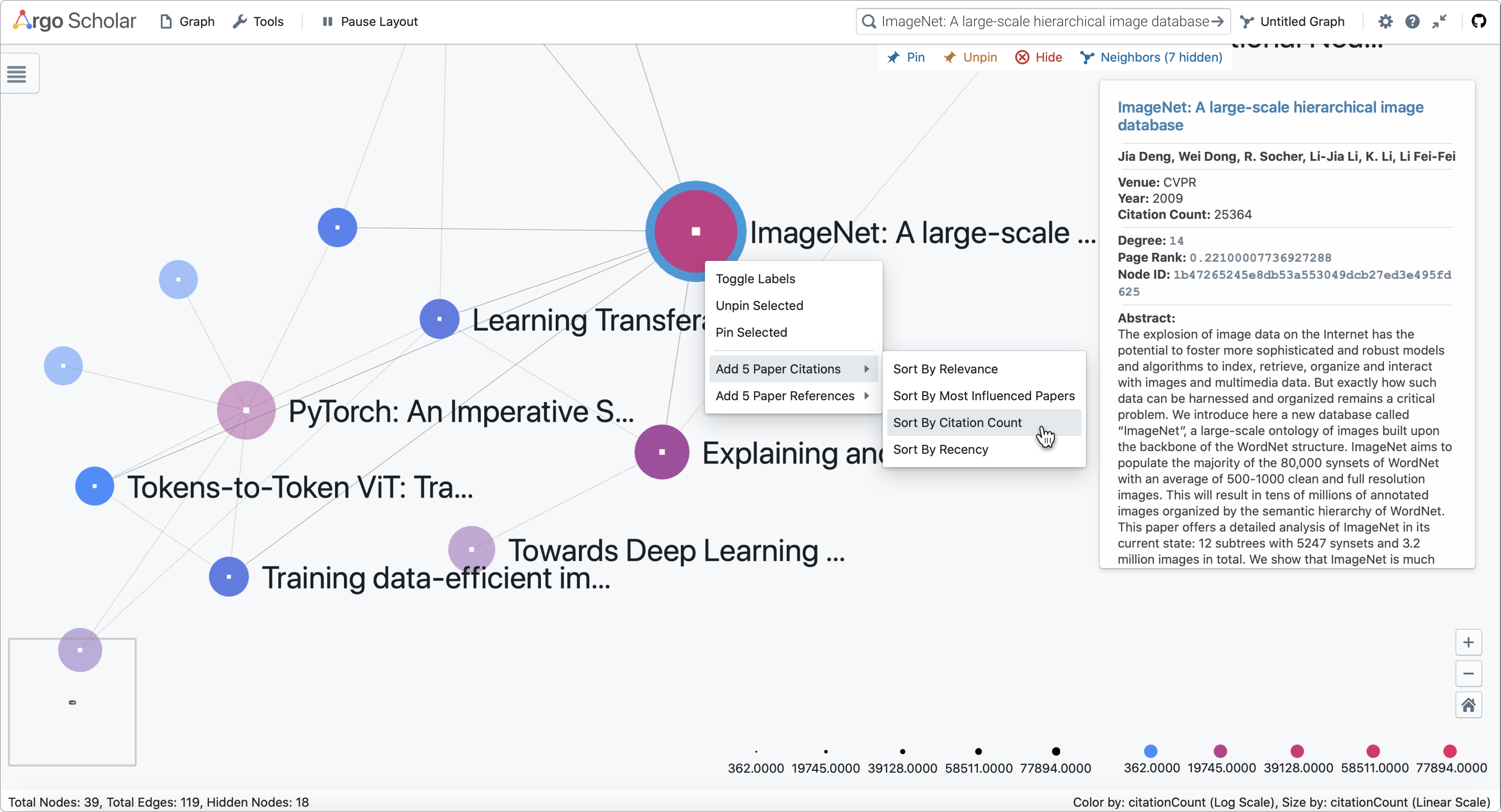}
  \caption{\tool{} visualizing a literature network of deep learning papers; nodes are papers and edges are citations. 
  }
  \label{fig:teaser}
\end{figure}

%% file: txt/Section-Background.tex
\section{Background: Visual Exploration of Literature with \tool{} }
\tool{}\cite{Li2021ArgoSI} is an open-source, web-based visualization tool for interactive exploration of literature and easy sharing of exploration results developed by the authors of this poster. \tool{} supports the method of incremental exploration, an effective technique to discover relevant and related work \cite{Sultanum2020UnderstandingAS, Wohlin2014GuidelinesFS}, on Semantic Scholar's live database of 200+ million papers \cite{Lo2020S2ORCTS}.

\tool{} (\autoref{fig:teaser}) enables users to generate personalized literature exploration results in real-time. \tool{}'s main view displays the user's collection of papers as a node-link diagram where papers are represented by nodes and are connected by citation edges. Users can easily expand their collection by querying for papers through keywords or adding the citations or references of existing papers based on relevance, citation count, influence, or recency \cite{Lo2020S2ORCTS}, and nodes can be dragged to be rearranged and display attributes adjusted to help the sensemaking process. Users can save or share a snapshot of their results as a URL on the cloud for free.

\tool{} was developed as an open-sourced web-based tool to help increase the availability and impact. Built with modern web technologies, such as Three.js WebGL rendering, \tool{} can smoothly render hundreds of papers and their relationships and is hosted at \textcolor{linkColor}{\url{https://poloclub.github.io/argo-scholar}}.
The open-sourced nature of \tool{} also differentiates itself from other existing tools.

%% file: txt/Section-Eval.tex
\section{Evaluation}
To investigate how users would interact with \tool{}'s features to explore literature networks and find relevant work and evaluate \tool{}'s effectiveness and usability, we conducted a large-scale user study of \participants{} participants. This section describes the study's design and findings.
\subsection{Experimental Design}

\hspace{\parindent}
\textbf{Participants.} \participants{} U.S.-based participants were recruited from Prolific\footnote{https://www.prolific.co/}, an online crowdsourcing platform designed specifically for academic research. Participants were screened to ensure they had at least one year of research experience, so they would be able to accurately gauge the effectiveness of \tool{} for finding related work. The resulting sample encompassed diverse knowledge backgrounds and research experiences (\autoref{tab:participants}). 76 participants were current students, while 46 were not currently enrolled. 21 listed a Doctorate degree as the high level of education completed or currently completing, 42 Master's, 44 Bachelor's, 12 GED or equivalent, and 3 others (JD, MD, and DDS). Notably, three were professors and two were postdoctoral researchers. 

\begin{table}[t]
  \caption{Participant background organized by National Science Foundation (NSF) research areas}.
  \label{tab:participants}
  \scriptsize%
	\centering%
  \begin{tabular}{l r}
  \toprule
   NSF Research Category & No.\\
  \midrule
    Social, Behavioral and Economic Sciences (SBE) & 51 \\ 
	Biological Sciences (BIO) & 24 \\
	Computer and Information Science and Engineering (CISE) & 21 \\ 
	Engineering (ENG) & 9 \\ 
	Mathematical and Physical Sciences (MPS) & 6 \\ 
    Education and Human Resources (EHR)	& 5 \\ 
    Integrative Activities (OIA) & 3 \\
    Environmental Research and Education (ERE) & 2 \\ 
    Geosciences (GEO) & 1 \\
  \bottomrule
  \end{tabular}%
  \vspace{-10pt}
\end{table}

\textbf{Procedure.}
After participants signed their consent forms electronically, they were provided with a background questionnaire and then a 1.5-minute tutorial of \tool{}, which described its visualizations and highlighted its features. In order to proceed with the study, participants were then given three tries to answer two simple multiple-choice pertaining to the tutorial video to ensure they had understood the instructions.
After that, participants were asked to use \tool{} on their computer's web browser for at least 10 minutes to build their own literature network on a topic of their choosing. After exploration, they were asked to submit their final results (must have included at least 10 relevant papers) and fill out a survey.
The study required no more than 20 minutes to complete, but many participants chose to explore the tool for longer. 
Each qualified participant was compensated \$4 with a bonus of \$2 for networks that demonstrated a clear effort to use the tool.

\subsection{Survey Results}
We measured the utility and usefulness of \tool{} using a 5-point Likert-scale (5 being \textit{Agree} and 1 being \textit{Disagree}). \autoref{tab:results} shows the average ratings for the 11 questions we asked. Overall, participants enjoyed using \tool{}, found it helpful at finding related work and understanding relationships between papers, and would like to use similar software for exploring literature.
They were particularly positive about \tool{}'s ability to help them find related work (4.48/5) and understand how papers are related to each other (4.30/5).
All key features of \tool{} were rated above a 4 for usefulness except for the \textit{Node Appearance Customization} panel, which might be due to certain customization rules already being preset. 

\subsection{Key Findings}
Below, we summarize the key finds distilled from the survey results and qualitative feedback and comments provided by our participants.

\medskip
\noindent
\textbf{Facilitating incremental exploration across disciplines.}
Our results reaffirm that incremental exploration works well for participants across a wide range of disciplines and helps them find related work and understand relationships between such articles.
For example, a participant pursing a biological sciences Doctorate student found \textit{``...it helped me to find some relevant papers I hadn't come across yet. This site could be very helpful for my dissertation.''}
This notion was supported by a speech-language pathology Master's student participant: \textit{``Everything about Argo Scholar makes the process so easy. I have never been able to map out citations and references for research papers so quickly in an organized fashion.''}

\medskip
\noindent
\textbf{No universal literature exploration tool.} The large-scale and diversity of our study provided many different viewpoints. While many found the node-link diagram representation helpful, such as a professor participant commending \textit{``...a nice tool to visualize the literature and I could see myself using this and showing it to graduate students to help them with their literature reviews.''}
and a Doctor of Medicine participant proclaiming \textit{``Love the node aspect and makes showing connections to other papers super easy!''}, some were not as receptive: \textit{``This type of visualization is really not how my mind works. I can see that it could be very powerful for some people, though''}. A learning curve for the system was also noted: \textit{``seemed a little difficult to use at first, but after a few minutes ... everything become more enjoyable''}.
It is encouraging the consensus found \tool{} useful and enjoyable; however, future work may focus on creating walk-up-and-use systems for literature exploration.

\medskip
\noindent
\textbf{Supporting increased user interactions.} 
Many participants expressed interest in features that would provide for more sensemaking opportunities. One participant wanted \textit{``to make my own ... lines to connect some of my own connections''}, a novel feature not found in citation network platforms.
Participants also asked for a \textit{``sticky note function''} prevalent in only a handful of tools \cite{ Sultanum2020UnderstandingAS}.
In general, users wanted more reign of their data: to annotate and adjust mental schemas to support the sensemaking process.
Comments such as \textit{``Very easy to publish ... Can save lots of time working with my colleagues''} along with the high ratings suggest sharing abilities would be valuable features for literature exploration tools.

\begin{table}[]
  \caption{Subjective ratings about \tool{} using 5-point Likert scale (5: Agree, 1: Disagree).}.
  \label{tab:results}
  \scriptsize%
	\centering%
  \begin{tabular}{l r}
  \toprule
   Question & Avg.\\
  \midrule
	I would like to use software like Argo Scholar in the future & 4.13 \\ 
	Helped me find related work & 4.48 \\ 
    Helped me understand how works were related to each other & 4.30 \\ 
    Usefulness of Graph View & 4.12 \\
    Usefulness of Exploration Dropdown & 4.64 \\
    Usefulness of Paper Information Panel & 4.58 \\
    Usefulness of Saving and Sharing Features & 4.37 \\
    Usefulness of Node Appearance Customizaation & 3.91 \\
    Was easy to use & 3.74 \\ 
	Was easy to understand & 3.70 \\
	I enjoyed using Argo Scholar & 4.11 \\ 
  \bottomrule
  \end{tabular}%
  \vspace{-10pt}
\end{table}

%% file: txt/Section-Conclusion.tex
\section{Conclusion and Discussion}
This paper presents a follow-up evaluation study of \tool{}, an open-sourced, web-based visualization tool that allows users to interactively and incrementally explore literature to gain insight and uncover relevant work.
From a large-scale observational study, we found \tool{} effective in helping users understand citational relationships within a field and uncover related work. We also condense lessons and suggestions from the feedback of our participants to provide inspiration for future work.